# The Dynamics of Platinum Precipitation in an Ion Exchange Membrane

S. F. Burlatsky[1], M. Gummalla[1*], V. V. Atrazhev[2], D. V. Dmitriev[2], N. Y. Kuzminyh[2], N. S. Erikhman[2]

[1]United Technologies Research Center, 411 Silver Lane, East Hartford, CT 06108, USA

[2]Russian Academy of Science, Institute of Biochemical Physics, Kosygin str. 4, Moscow, 119334, Russia

## Abstract

Microscopy of polymer electrolyte membranes that have undergone operation under fuel cell conditions, have revealed a well defined band of platinum in the membrane. Here, we propose a physics based model that captures the mechanism of platinum precipitation in the polymer electrolyte membrane. While platinum is observed throughout the membrane, the preferential growth of platinum at the band of platinum is dependent on the electrochemical potential distribution in the membrane.

In this paper, the location of the platinum band is calculated as a function of the gas concentration at the cathode and anode, gas diffusion coefficients and solubility constants of the gases in the membrane, which are functions of relative humidity. Under $H_2$/$N_2$ conditions the platinum band is located near the cathode-membrane interface, as the oxygen concentration in the cathode gas stream increases and/or the hydrogen concentration in the anode gas stream decreases, the band moves towards the anode.

The model developed in this paper agrees with the set of experimental data on the platinum band location and the platinum particle distribution and size.

* Corresponding author; 411 Silver Lane, MS 129-13, United Technologies Research Center, East Hartford, CT 06108; Fax: 860 660 8240; Email: gummalm@utrc.utc.com

## Introduction

Polymer electrolyte membranes (PEM) are critical components of the fuel cell that enables conduction of protons while preventing the flow of electrons through it. Additionally, the membrane separates the fuel and the oxidant streams that react electrochemically at the anode and cathode. However, the membranes are not perfect separators and there is some leakage of the gas (hydrogen from the anode and oxygen from the cathode) across the membrane called cross over. As discussed later in the paper, the cross over gas fluxes determine catalyst particle precipitation in the membrane.

Platinum is a commonly used catalyst in the fuel cell electrodes. While its activity is high, the stability [1-4] under fuel cell conditions limits its durability [5-7]. Instability of Pt at high potentials is greater and is aggravated by potential cycling [1, 2]. While the impact of dissolution of platinum on electrochemical surface area loss is extensively discussed [1, 2, 4, 8] precipitation of the platinum and re-distribution within the membrane electrode assembly is less known [9-14]. The subsequent possibility of such platinum particles to generate radicals within the membrane is relatively new [5-7]. These radicals tend to degrade the polymer resulting in its failure due to high gas cross-over. Therefore, fundamental understanding of the mechanism of Pt precipitation is important to develop mitigation strategies and enhance the life of the fuel cell.

Patterson et al, [9] have shown the existence of a Pt band in the post test analysis of the membrane that underwent $H_2$/air cycling. The location of the band of Pt is a function of the operating conditions [11, 13]. Yasuda et al, [10] observed that Pt band is formed at the cathode membrane interface after $H_2/N_2$ potential cycling. Additionally, they identified varied morphological features of the Pt precipitates in the membrane resulting from varied operational conditions or membrane thickness. A semi-qualitative model to predict the location of the band is also discussed in [13]. The Pt band location is determined by the coordinate $X_0$ representing the abrupt drop of potential [15]. Transmission electron microscopy (TEM) analysis of the membrane after $H_2$/Air and $H_2/O_2$ cycles is shown in Figure 1. Figure 1 indicates the size of the Pt particles in the membrane as a function of location. Figure 1a shows tiny Pt particles throughout the membrane and large particles near the $X_0$ (indicated by a vertical line in Fig. 1), for $H_2$/Air cycled membrane. Figure 1b shows a similar pattern observed in a set of experiments when membrane was cycled under hydrogen/oxygen conditions. However, in this case, the Pt band is located in the middle of the membrane. TEM analysis of the cycled membrane indicates that the location of the Pt band varies with the cathode oxygen concentration. These observations are in qualitative agreement with the membrane potential model presented in [15].

Potential of the membrane can be interpreted as the potential of the proton or of the catalyst particle in the membrane. Here, we follow the approach of [16], in which the local membrane potential was attributed to the potential of a small size platinum probe inserted at a particular location in the membrane vs. standard hydrogen electrode (SHE).

This potential is controlled by concentration of oxygen and hydrogen at the probe surface. As discussed in [15] the oxygen reduction reaction (ORR) and hydrogen oxidation reaction (HOR) at the surface of the probe are diffusion controlled when the probe is large and kinetic controlled when the probe is small. Furthermore, the probe significantly changes the local gas concentrations in its close proximity, with no significant change in the membrane bulk. The probe is preferentially exposed to hydrogen if inserted between the anode and $X_0$ and it is preferentially exposed to oxygen if inserted between the cathode and $X_0$. The local gas concentration defines the typical shape of the potential distribution in the membrane, shown in Fig. 2. From the perspective of Pt precipitation, the local potential is important because it determines the rate of nucleation, electrochemical dissolution and precipitation of platinum at that location.

The focus of the current paper is to provide the fundamental mechanism of Pt precipitation and growth and a mathematical model that captures platinum diffusion, precipitation and band formation in the membrane. Subsequently, the band location and platinum particle distribution observed experimentally [10] is explained. In the first part of the paper we describe the model and simulation method followed by the discussion of the results and comparison with experimental data.

**Model Description**

The model description part is divided into 3 sections. The first section qualitatively describes the electrochemical/chemical mechanisms of platinum nucleation and growth in the membrane. The second section provides the quantitative model of platinum nucleation stage. The third section provides the model of Pt growth stage with mass conservation equations for oxygen, hydrogen, platinum particles, platinum ions and the local charge.

**SECTION I: Electrochemical/chemical mechanism of platinum nucleation and growth**

The mechanism of platinum precipitation is schematically shown in Fig. 3. Platinum dissolves from the cathode at high potential, diffuses into the membrane in ionic state, and forms neutral nuclei that grow into particles depending on the local membrane potential. The dissolution equation is

$$Pt \rightarrow Pt^{++} + 2e^{-} \tag{1}$$

The Pt ions diffuse into the membrane driven by the concentration gradient and precipitate in the presence of hydrogen cross over forming Pt atoms, $Pt^{0}$.

$$Pt^{++} + H_2 \rightarrow Pt^{0} + 2H^{+} \tag{2}$$

The Pt atoms are assumed to diffuse and agglomerate to form Pt particles, $(Pt)_n$, to minimize their high surface energy.

$$(Pt)_n + Pt^{0} \rightarrow (Pt)_{n+1} \tag{3}$$

Additionally, Pt ions precipitate electrochemically at low potentials on platinum particles formed in the membrane through the reactions,

$$\begin{aligned} &(Pt)_n + Pt^{++} \rightarrow (Pt)_{n+1}^{++} \\ &(Pt)_{n+1}^{++} + H_2 \rightarrow (Pt)_{n+1} + 2H^{+} \end{aligned} \tag{4}$$

The dynamics of platinum precipitation and growth is defined by the local potential and Pt particle size. This dynamics can be virtually divided into two stages: initial nucleation of Pt particles in membrane and Pt particles growth governed by electrochemical kinetics.

The size of Pt particles at initial nucleation stage is smaller than the mean size of water clusters in hydrated Nafion membrane (~2.5nm). So, Pt particles can easily diffuse in the membrane bulk with diffusion coefficient of the same order of magnitude as that in water. Thus, taking into account high mobility of small Pt particles platinum precipitates homogeneously throughout the membrane, in this stage. This results in uniform Pt nucleus concentration (defined as the number of Pt particles per unit volume) in membrane.

At initial nucleation the platinum particles are very small, ≤2.5nm, and are not electrochemically active due to reasons described in appendix B. Therefore, electrochemical reactions in the membrane are very slow and, according to [15], the potential of Pt particles is almost constant across the membrane.

Initial nucleation stage completes and particle growth stage starts when the platinum particles become larger than the mean size of water clusters in the membrane ~2.5nm, the platinum particles cease to diffuse and become immobile. That results in non-uniform growth of immobile Pt particles in the vicinity of $X_0$. Therefore, despite the uniform distribution of Pt particle concentration, the platinum mass distribution is not uniform.

A one dimensional model along the thickness of the membrane is developed to capture the mechanism of platinum dissolution, precipitation, and growth within the membrane. The reaction and diffusion equations for $O_2$, $H_2$, Pt atoms, Pt ions and Pt particles are derived and solved numerically. For simplicity, the model assumes a constant flux of platinum ions into the membrane from the cathode. This also enables a decoupling of the well discussed [1, 2] electrode platinum dissolution and precipitation dynamics.

The average Pt ion flux is estimated based on the ratio of experimentally observed total platinum mass accumulated in the membrane to the total cycling time. For the potential cycling conditions considered in this article the average flux is estimated to be,

$$J_{Pt,total} = 10^{-13}\left[\frac{mol}{cm^2 \cdot s}\right]. \tag{5}$$

The evolution of platinum ions in a pristine membrane is divided into two stages, as described next.

**SECTION II: Initial platinum nucleation in the membrane**

During this stage the platinum ions from the cathode react with hydrogen forming atomic Pt, as indicated by Eq. (2). Atomic Pt diffuses fast to agglomerate into Pt particles. In this stage, the hydrogen concentration in the membrane, due to cross over, is much higher than the Pt particle concentration. Therefore, Pt ions are reduced to Pt atoms more preferably. As discussed in Appendix B electrochemical reactions rates are suppressed at this stage. The schematic representation of the hydrogen and oxygen concentration profile is shown in Fig. 4a. The concentration profiles are linear by virtue of the diffusion equation with no appreciable reaction between them. The equation for Pt ion diffusion and reduction by hydrogen in the membrane is

$$-D_{Pt^{++}}\frac{d^2C_{Pt^{++}}}{dx^2} = -K_{Pt^{++}H_2} \tag{6}$$

Here, x is the coordinate across membrane, $D_{Pt++}$~$10^{-6}$cm$^2$/s is diffusion coefficient of Pt ion in the membrane and $K_{Pt^{++}H_2}$ is the rate of Pt ion reduction reaction (2). Reaction

(2) is assumed to be diffusion controlled and, therefore, very fast. Under this assumption, most Pt ions are reduced by hydrogen on the cathode side of membrane and only a very small fraction can migrate to the anode. The atomic Pt is assumed to have same diffusion coefficient, $D_{Pt}$, as that of the ionic platinum. The typical diffusion time of Pt atom $\tau_{diff}$ in the membrane is

$$\tau_{diff} = \frac{L_m^2}{D_{Pt}} \tag{7}$$

For membrane thickness $L_m$ of 20μm and $D_{Pt}$ of $10^{-6}$cm$^2$/s, the $\tau_{diff}$ is 4s. Since this time is much shorter than the time scale to observe initial Pt precipitation, a uniform distribution of Pt particles in the membrane is highly likely. The aggregation process of the small Pt particles is assumed to be diffusion limited. The diffusion controlled aggregation rate constant, $K_{agg}$, of two particles of radius $R_n$ and $R_m$ is

$$K_{agg}(m,n) = 4\pi(D_n + D_m)(R_n + R_m) \tag{8}$$

Assuming the Stokes-Einstein equation for the mobility of platinum particle in Nafion, the diffusion coefficient is inversely proportional to the particle size,

$$D_n \approx \frac{D_{Pt} a_{Pt}}{R_n}. \tag{9}$$

Here, $a_{Pt} = 1.4\,\mathring{A}$ is the radius of Pt atom.

Combining Eqs. 8 and 9

$$K_{agg}(m,n) = 4\pi D_{Pt} a_{Pt} \frac{R_n}{R_m}\left(1 + \frac{R_m}{R_n}\right)^2. \tag{10}$$

Assuming $R_n \approx R_m$, the aggregation rate constant is independent of Pt particle radius, and $K_{agg} \approx 10^{-12}\, cm^3/s$.

The aggregation rate of small and large particles would be higher than that between particles of equal size. However, the size effect substantially complicates the calculations without appreciable change in results. Hence, the slowest aggregation rate constant is assumed.

The dynamic concentration of platinum is given by standard equation of diffusion-controlled agglomeration [17]

$$\frac{dc_k}{dt} = \frac{K_{agg}}{2} \sum_{m+n=k} c_m c_n - K_{agg} c_k \sum_{m=1}^{\infty} c_m + J\delta_{k=1} . \qquad (11)$$

Here, $c_k$ is the concentration of Pt particles containing $k$ Pt atoms. On the right hand side of Eq. (11) the first term corresponds to all combinations of Pt particles that result in a concentration $c_k$, and the second term is loss of $c_k$ due to agglomeration with all other size platinum particles. The third term is a source of Pt atoms in membrane, which is equal to a mean volumetric source of Pt ions,

$$J = \frac{J_{Pt,total} N_A}{L_m} , \qquad (12)$$

estimated at $J \approx 10^{13} cm^{-3} s^{-1}$ with $N_A$ the Avogadro number.

Initial condition for the aggregation equation assumes no Pt particles in the membrane at t=0:

$$c_k(0) = 0 \qquad (13)$$

Equation (11) is the aggregation equation with constant kernel, which is solved analytically using the generating function formalism [17] and mathematical details are not discussed here. Solution of Eq. (11) with the initial condition Eq. (13) results in total concentration of Pt particles $n_{Pt}(t) \equiv \sum_k c_k(t)$ increasing with time as:

$$n_{Pt}(t) = n_{Pt}^{0} \tanh \frac{t}{\tau_{agg}}, \qquad (14)$$

where,

$$n_{Pt}^{o} = \sqrt{\frac{2J}{K_{agg}}}, \qquad (15)$$

and

$$\tau_{agg} = \sqrt{\frac{2}{JK_{agg}}}. \qquad (16)$$

Here, $n_{Pt}^{o} \approx 10^{13} cm^{-3}$ is characteristic Pt particle concentration and $\tau_{agg} \approx 1s$ is characteristic aggregation time.

Interestingly, the number of Pt particles in the membrane approaches an asymptotic constant value, per Eq (14). This is the key result of initial nucleation stage and the number of Pt particles remains constant through the next stage of Pt particles growth, as discussed in the next section.

The evolution of mean Pt particle size, r, is governed by mass balance equation

$$n_{Pt}\left\langle r^{3}\right\rangle = a_{Pt}^{3} \cdot J \cdot t \qquad (17)$$

As follows from Eq. (17) the mean Pt particle size grows as $t^{1/3}$ and grows to a size r~2.5nm after t~3hr, which defines the duration of initial stage. Pt particles concentration $n_{Pt}^{o}$ is used in numerical model of Pt particles growth.

**SECTION III: Pt particles growth in membrane**

When the mean size of Pt particles exceeds the critical radius $r_{c,el}$*~2.5nm* (see Appendix B for details) the electrochemical reactions (ORR, HOR and Pt dissolution) rates on Pt particle surfaces become non trivial. This changes the distribution of gases in the membrane, and the concentration profiles of hydrogen and oxygen are schematically shown in Fig. 4b. Approximately at the same time Pt particles become immobile when they reach the mean size of the water clusters in hydrated Nafion membrane, $r_c$~2.5nm. The growth of Pt particles occurs as follows. Pt ions freely diffuse from the cathode to the plane $X_0$, where they turn into atomic platinum due to excess hydrogen. These neutral Pt atoms aggregate with Pt particles resulting in maximum growth of Pt particles near $X_0$.

While there is ability for Pt atoms generated due to reaction (2) to diffuse in the membrane and aggregate to form new small Pt particles, the probability for the newborn Pt atom to grow to the size $r_c$~2.5nm without any collision with the immobile large Pt particles is negligible. Therefore, the concentration of Pt particles is assumed to be constant with time and uniform in space. Since the distribution of the Pt ions and Pt atoms in the membrane is non-uniform, the growth of immobile Pt particles becomes non-uniform, resulting in non-uniform platinum mass distribution in the membrane.

The Pt particle size distribution in the membrane is spatially and time varying, thus represented as *r(x,t).* The platinum evolution is highly nonlinear and coupled with the reactant concentration, its own size and potentials described by the following series of equations. The model accounts for four species, namely $O_2$, $H_2$, Pt ion, Pt atom and Pt particles of radius r. These are treated as distributed variables along the thickness of the membrane. The model equations with boundary conditions are summarized in Table 1. Equations for the reaction rates incorporated in the model are summarized in Table 2.

## Oxygen Transport

The diffusion-reaction equation for oxygen is

$$D_{O_2}\frac{\partial^2 C_{O_2}}{\partial x^2}=K_{O_2}\,. \qquad (18)$$

Here, $D_{O_2}$ is the diffusion coefficient of oxygen in the membrane, $C_{O_2}$ is the concentration of oxygen in the membrane. $K_{O_2}$ is the rate of ORR

$$\mathrm{O}_2+4\mathrm{H}^+ +4\mathrm{e}^- \leftrightarrow 2\mathrm{H}_2O\,. \qquad (19)$$

The ORR rate is proportional to the local surface area of the catalyst $4\pi r^2 n_{Pt}$ and current of ORR, $i_{O_2}$

$$K_{O_2}(x,t)=\frac{4\pi\, r^2(x,t)n_{Pt}}{4F}i_{O_2}\,. \qquad (20)$$

Here, F=96485 C/mol is Faraday's constant. The current of ORR $i_{O_2}$ is calculated by Buttler-Volmer equation with diffusion limitation (see ref. [15] for details)

$$i_{O_2}=i^0_{O_2}\frac{\dfrac{C_{O_2}}{H_{O_2}C^{ref}_{gas}}\exp\left(\dfrac{U_0-\Phi}{b}\right)-\exp\left(-\dfrac{U_0-\Phi}{b}\right)}{1+\dfrac{i^0_{O_2}}{i_{\lim,O_2}}\exp\left(\dfrac{U_0-\Phi}{b}\right)}\,. \qquad (21)$$

Here, $H_{O2}$ is dimensionless Henry constant for oxygen in the membrane, $i^0_{O_2}=10^{-8}A/cm^2$ [2] is exchange current, $U_0=1.23V$ is equilibrium potential of ORR at normal conditions, $C^{ref}_{gas}=\dfrac{P_{atm}}{RT}$ is a reference gas concentration, $\Phi$ is the local potential, and $b$ is Tafel constant defined as $b=\dfrac{RT}{F}\approx 0.028V$. The ORR limiting current, $i_{\lim,O_2}$, accounts for the local gas diffusion limitation on Pt particle, given by the formula

$$i_{\lim,O_2}=4FC^{ref}_{gas}\frac{H_{O_2}D_{O_2}}{r}\,. \qquad (22)$$

Where, R=8.31 J/mol/K is universal gas constant.

The boundary conditions for the oxygen transport Eq. (18) are: zero concentration of $O_2$ at the Anode/Membrane Interface (AMI) due to very fast oxygen reduction at anode potential

$$C_{O_2}\Big|_{AMI} = 0, \tag{23}$$

and at the Cathode/Membrane Interface (CMI)

$$C_{O_2}\Big|_{CMI} = H_{O_2} C_{O_2}^{chan}. \tag{24}$$

Where, $C_{O_2}^{chan}$ is oxygen concentration in cathode gas channels of the fuel cell.

## Hydrogen Transport

Analogous to oxygen, diffusion-reaction equation for hydrogen is

$$D_{H_2} \frac{\partial^2 C_{H_2}}{\partial x^2} = K_{H_2} + K_{Pt^{++}H_2}. \tag{25}$$

Here, $D_{H_2}$ is the diffusion coefficient of hydrogen in the membrane, $C_{H_2}$ is concentration of hydrogen in the membrane. $K_{Pt^{++}H_2}$ is the rate of Pt ion reduction by hydrogen, reaction (2). This reaction is assumed to be diffusion controlled and the reaction rate is determined by equation

$$K_{Pt^{++}H_2} = 4\pi D_{H_2} a_{Pt} N_A C_{H_2} C_{Pt^{++}}. \tag{26}$$

$K_{H_2}$ is the rate of HOR,

$$\mathrm{H_2} \leftrightarrow 2\mathrm{H^+} + 2\mathrm{e^-}. \tag{27}$$

Similar to ORR the HOR rate, $K_{H_2}$, is proportional to the local surface area of the catalyst $4\pi r^2 n_{Pt}$ and to the current of HOR $i_{H_2}$,

$$K_{H_2}(x,t) = \frac{4\pi r^2(x,t) n_{Pt}}{2F} i_{H_2}. \tag{28}$$

$i_{H_2}$ is proportional to the local hydrogen concentration and is given by

$$i_{H_2} = i_{H_2}^0 \frac{\frac{C_{H_2}}{H_{H_2} C_{gas}^{ref}} \exp\left(\frac{\Phi}{b}\right) - \exp\left(-\frac{\Phi}{b}\right)}{1 + \frac{i_{H_2}^0}{i_{\lim,H_2}} \exp\left(\frac{\Phi}{b}\right)} . \qquad (29)$$

Here, $H_{H2}$ is dimensionless Henry constant for hydrogen in the membrane, $i_{H_2}^0 = 0.01 A/cm^2$ [2] and the diffusion limited current is

$$i_{\lim,H_2} = 2FC_{gas}^{ref} \frac{H_{H_2} D_{H_2}}{r} . \qquad (30)$$

The boundary conditions for hydrogen transport Eq. (25) are similar to oxygen case: zero concentration of $H_2$, at the CMI due to very fast hydrogen oxidation at cathode potential

$$C_{H_2}\Big|_{CMI} = 0 , \qquad (31)$$

and at AMI

$$C_{H_2}\Big|_{AMI} = H_{H_2} C_{H_2}^{chan} . \qquad (32)$$

Here, $C_{H_2}^{chan}$ is hydrogen concentration in anode gas channels of the fuel cell.

## Pt ion transport

The diffusion-reaction equation for Pt ions is

$$D_{Pt^{++}} \frac{\partial^2 C_{Pt^{++}}}{\partial x^2} = -K_{Pt^{++}} + K_{Pt^{++}H_2} . \qquad (33)$$

Here, $K_{Pt^{++}}$ is reaction rate of electrochemical Pt dissolution/precipitation. This reaction rate is a function of potential, $\Phi$, Pt ion concentration in the membrane, $C_{Pt^{++}}$ , and the radius of the Pt particle *r*. The reaction rate is proportional to the surface area of Pt, given by [18]

$$K_{Pt^{++}} = 4\pi r^2 n_{Pt} k_1 \frac{\exp\left(2\frac{\alpha_1(\Phi-U_1)}{b}\right) - \frac{C_{Pt^{++}}}{C_{ion}^{ref}}\exp\left(-2\frac{(1-\alpha_1)(\Phi-U_1)}{b}\right)}{1+\frac{k_1}{k_{\lim Pt^{++}}}\exp\left(-2\frac{(1-\alpha_1)(\Phi-U_1)}{b}\right)}. \qquad (34)$$

Here, $k_1$=1.8·10$^{-14}$mol/cm$^2$s is the dissolution rate constant, $\alpha_1$=0.3, $U_1$=1.188V is the equilibrium potential of Pt dissolution, $C_{ion}^{ref} = 1\ mol/l$ is a reference ion concentration. The Pt precipitation diffusion rate $k_{\lim Pt^{++}}$ in Eq. (34) accounts for the local diffusion limitations on Pt particle and is given by equation:

$$k_{\lim Pt^{++}}(x) = C_{ion}^{ref}\frac{D_{Pt^{++}}}{r} \qquad (35)$$

Pt ions flow into the membrane through CMI, the rate of the Pt ion flow determines the boundary condition to Eq. (33) at CMI:

$$-D_{Pt^{++}}\frac{dC_{Pt^{++}}}{dx}\bigg|_{CMI} = J_{Pt,total} \qquad (36)$$

Pt ions that diffuse up to the AMI immediately precipitate on the anode due to low anode potential. Thus, the boundary condition for Pt ions at AMI is

$$C_{Pt^{++}}\big|_{AMI} = 0. \qquad (37)$$

## Pt atom transport

The diffusion-reaction equation for Pt atoms is

$$D_{Pt}\frac{\partial^2 C_{Pt}}{\partial x^2} = -K_{Pt^{++}H_2} + K_{Pt} \qquad (38)$$

$K_{Pt}$ is the aggregation rate of neutral Pt atoms with Pt particles. The reaction rate of aggregation of Pt atoms with Pt particles is also assumed to be diffusion controlled, and proportional to the Pt particle surface area

$$K_{Pt} = 4\pi D_{Pt} r n_{Pt} C_{Pt} \,. \tag{39}$$

Furthermore, Pt atoms at the membrane interface with anode or cathode are assumed to deposit on the electrode. Thus the boundary conditions for Eq. (38) at CMI and AMI are

$$C_{Pt}\big|_{CMI} = 0 \text{, and} \tag{40}$$

$$C_{Pt}\big|_{AMI} = 0 \,. \tag{41}$$

## Evolution of Pt particle radius

Pt particles are characterized by their mean radius r(x,t), and the dynamic evolution is given by

$$\frac{4\pi r^2 n_{Pt}}{v_{Pt}} \frac{\partial r(x,t)}{\partial t} = K_{Pt} - K_{Pt^{++}} \,. \tag{42}$$

Here, $v_{Pt}$=9.286 cm$^3$/mol is molar volume of crystalline Pt. At the beginning of the growth stage Pt particles are of uniform size of $r_c$*=2.5nm* and are distributed uniformly in the membrane. Thus, the initial condition for Eq. (42) is

$$r(x) = r_c \tag{43}$$

## Local electrical charge conservation

Pt particles in the membrane are disconnected with the external circuit and the Pt particle potential (relative to electrolyte potential) Φ(x) is determined by charge conservation equation for disconnected Pt particles in the membrane. There are three electrochemical reactions taking place on the Pt particle surface: Pt dissolution/precipitation, ORR and HOR. The local electric charge conservation law is

$$2K_{O_2} - K_{H_2} - K_{Pt^{++}} = 0\,. \tag{44}$$

This equation determines the distribution of the Pt particle potential in the membrane. With the assumption that Pt dissolution rate is much lower than ORR and HOR in the membrane, $K_{Pt^{++}} << K_{O_2}, K_{H_2}$, Eq. (44) is further simplified.

Additionally, typical diffusion time for all species $\tau_\alpha = \frac{L_m^2}{D_\alpha} \sim 1s$ is much less than the characteristic time, τ, of the redistribution of given species in the membrane, which is of the order of the time of experiment (several hours). Therefore, we neglected the terms $\frac{\partial C_\alpha}{\partial t}$ in the left-hand sides of all the transport equations, except for the growth of Pt particles.

## Numerical Approach

The mathematical equations for Pt particles growth stage are implemented and solved using a commercial software, gPROMS. Equations for Pt particle nucleation are solved analytically and $n_{Pt}$ is computed per Eq. 17. This value is constant for the stage of Pt particles growth. Initial radius for growth stage is assumed to be $r_c$ = 2.5nm. In this stage, equations for oxygen, hydrogen, Pt ions and Pt atoms concentrations, and potential $\Phi(x)$ of the Pt particle within membrane are solved numerically. The net growth of Pt mass in the membrane is determined by the integral of the distributed sink/source terms of Pt ions and Pt atoms in the membrane and that lost or gained at the cathode/membrane and membrane/anode interface. This aggregated mass of platinum is distributed evenly to all particles to determine the radius of typical Pt particle. The set of differential and algebraic equations are solved as a time dependent problem. Central finite difference scheme is used with, typically, 100 nodes.

## Results and Discussion

The evolution of the concentration profiles of hydrogen, oxygen, mean Pt particle size in the membrane and potential of the membrane are strongly coupled through equation (44), and all dependent equations thereof. The simulation results of cycling a fuel cell from 0.95 V to 0.6 V, with hydrogen on the anode and pure oxygen on the cathode at short time, intermediate time and long time, relative to the start of potential cycling are discussed next. For this simulation, Eq. 5 is used as the platinum ion flux into the membrane.

At time t=0.1 hr, Pt particles are small (r<1nm) and their fast diffusion leads to uniform distribution in the membrane. Since these small Pt particles are not sufficiently chemically or electrochemically active, there is no consumption of oxygen and hydrogen in the membrane resulting in linear concentration profiles for $H_2$ and $O_2$, as shown in Fig. 5a. The potential near both electrode interfaces with the membrane exhibits sharp spatial gradient because of the logarithmic dependence on gas concentration through Nernst equation. In the bulk of the membrane, the potential lies between the anode-membrane interface potential and the cathode-membrane interface potential. A small kink in potential is observed near $X_0$ even at this stage of deposition due to apparent activity of platinum. As discussed in [15] the size of platinum probe determines if the HOR/ORR are diffusion controlled or kinetically controlled. A simplified criterion for the onset of diffusion controlled reactions as a function of platinum particle size is provided in Appendix C. A key difference between platinum probe in [15] and the current system is the presence of multiple platinum particles. It is to be noted that in this stage, the platinum particles are very small to have stable electrochemical reactions, and the potential is treated as “apparent potential”, assuming stable reactions.

At the size greater than 2.5nm, the platinum particles can support significant HOR and ORR. Hydrogen and oxygen are consumed in the membrane mostly near $X_0$, which is defined by stoichiometric relation between hydrogen and oxygen flux to make water $2J_{O_2}(X_0) = J_{H_2}(X_0)$, a consequence of charge conservation law (44). The potential profile is changed because the local reaction rate at the individual particles becomes comparable to the local diffusion limit, see Fig. 5b (t = 10 hrs). When sufficient platinum is accumulated in the membrane the reactant profiles are less linear and more concave, the conditions when this change occurs is discussed in Appendix D.

Approximately at this time Pt particles are so large (>2.5nm) that they are immobile in the membrane matrix and the platinum mass distribution becomes non-uniform. Large platinum particles are also electrochemically very active, resulting in greater consumption of the crossover hydrogen and oxygen close to the $X_0$ plane. This results in very low concentration of hydrogen on the cathode side of $X_0$ plane and low concentration of oxygen on the anode side of $X_0$ plane. Therefore, Pt ions diffuse from the cathode to the $X_0$ plane, where they encounter and react with hydrogen forming Pt atoms. Subsequently, neutral Pt atoms meet and aggregate with neighboring Pt particles. Therefore, the fastest growth of immobile Pt particles takes place near the $X_0$ plane, where the Pt atoms are preferentially generated. This fact is confirmed by numerical simulation shown in Fig. 5c (t = 100 hrs). A slight asymmetry of Pt particle size distribution with respect to the $X_0$ plane is caused by Pt dissolution reaction, which occurs at high potential on the cathode side of the $X_0$ plane.

With this fundamental understanding of the evolution of platinum in the membrane, we discuss the effects of potential cycling on hydrogen air and hydrogen nitrogen systems. The most distinct effect of change in reactants or its concentration is the location of Pt precipitation. Figure 6a is a SEM image showing the band of platinum in the membrane after 20,000 cycles under $H_2$/air conditions. The corresponding simulation results of the potential and Pt radius distribution in the membrane are shown in Figures 6b and 6c. The physics is very similar to that described for the $H_2/O_2$ case, with the location of the $X_0$ plane shifting to the cathode due to the lower oxygen concentrations. The results shown here are in agreement with the TEM images from Figure 1, where the large particles are found nearer to the cathode in the $H_2$/air case and in the middle of the membrane in the $H_2/O_2$ case.

Next, we discuss platinum precipitation in hydrogen nitrogen systems, with nitrogen on the cathode side. In this case, the cathode potential is cycled by an external power source causing Pt dissolution at controlled humidity. There is no oxygen reduction reaction at the Pt particles in the membrane, due to absence of the oxygen, the hydrogen oxidation in the membrane is also absent, for charge neutrality. Thus the potential of membrane is equal to hydrogen equilibrium potential (~0 V) and hydrogen crosses over without consumption in the membrane. This results in the $X_0$ plane appearing at the interface of cathode and membrane. Pt ions from the cathode diffuse to the membrane and react with hydrogen at cathode/membrane interface reducing to Pt atoms by reaction (2), and they subsequently agglomerate into particles. Therefore, this leads to preferable precipitation of Pt in $H_2/N_2$ conditions at the cathode/membrane interface. This is in agreement with the TEM images of Yasuda et al [10], where they show a pronounced Pt band at cathode/membrane interface under $H_2/N_2$ conditions after potential cycling.

## Summary

A physics-based mathematical model that predicts the time dependent evolution of platinum in the membrane, under fuel cell conditions, has been developed. The model describes transport of oxygen, hydrogen, Pt ions, Pt atoms and Pt particles in the membrane. It takes into account electrochemical reactions (ORR, HOR, platinum dissolution/precipitation) as well as Pt ions reduction by hydrogen and platinum agglomeration reactions, to predict the experimentally observed platinum band in the membrane.

The model predicts that Pt particle concentration in the membrane is determined at very early stage by platinum aggregation rate and Pt ion flux from the cathode and does not change. Additionally, Pt particle concentration is uniformly distributed in the membrane due to high mobility of small Pt particles at early stages. In spite of uniform Pt particle concentration, the Pt mass distribution becomes non-uniform after Pt particles reach the critical size ~2.5nm (typical size of water cluster in Nafion) and become immobile in the membrane matrix.

When there is sufficient amount of platinum in the membrane most of the hydrogen and oxygen are consumed by ORR and HOR on Pt particle surfaces near the $X_0$ plane, which is determined by the location where the oxygen and hydrogen fluxes in the membrane are in stoichiometry [15] for water production.

The key prediction is that the platinum band is located near the $X_0$ plane, which is strongly dependent on the potential distribution in the membrane. This is determined by the oxygen and hydrogen concentrations in the cathode and anode, respectively.

Therefore, the location of platinum band shifts with the change of inlet oxygen and/or hydrogen concentrations. In particular, lowering of the oxygen concentration (with fixed hydrogen concentration) leads to the shift of the platinum band to the cathode. In the limit of zero oxygen concentration (hydrogen-nitrogen system) the platinum band is located at the cathode membrane interface. These theoretical predictions of the model are in very good agreement with the experimental data of potential cycling on hydrogen-air and hydrogen-nitrogen systems, reported in literature [10].

**Acknowledgements**

The authors would like to thank Dr. Jarvi, Dr. Condit, and Dr. Madden for helpful discussions, Dr. Weiss and Dr. Snow for providing the TEM pictures. UTCPower and UTRC assisted in meeting the publication cost of this article.

APPENDIX A

| | |
|---|---|
| x | dimensionless, coordinate across membrane normalized by membrane thickness |
| $\alpha_1$ | dimensionless, symmetry constant for Pt dissolution reaction |
| $a_{Pt}$ | 1.4 Angstroms, radius of Pt atom |
| $b$ | $V$, Tafel slope |
| $C_{ref}$ | $= 3.6 \cdot 10^{-5}\ mol/cm^3$ , gas concentration at 100 kPa, 60 $^0$C |
| $c_{H_2}$ | dimensionless, hydrogen concentration at particle surface normalized by $C_{ref}$ |
| $c_{O_2}$ | dimensionless, oxygen concentration at particle surface normalized by $C_{ref}$ |
| $c_{H_2}^{(0)}$ | dimensionless, hydrogen concentrations at the anode normalized by $C_{ref}$ |
| $c_{O_2}^{(0)}$ | dimensionless, oxygen concentrations at cathode normalized by $C_{ref}$ |
| $c_{H_2}^{(b)}$ | dimensionless, hydrogen concentration normalized by $C_{ref}$ in membrane bulk |
| $c_{O_2}^{(b)}$ | dimensionless, oxygen concentration normalized by $C_{ref}$ in membrane bulk |
| $C_{ref}$ | $mol.cm^{-3}$, reference gas concentration |
| $C_{Pt^{++}}^{ref}$ | $mol.cm^{-3}$, reference concentration of ionic species |
| $D_{H_2}$ | $cm^2 \cdot s^{-1}$, diffusion coefficient of hydrogen in the membrane |
| $D_{O_2}$ | $cm^2 \cdot s^{-1}$, diffusion coefficient of oxygen in the membrane |
| $D_{Pt^{++}}$ | $cm^2 \cdot s^{-1}$, diffusion coefficient of Pt ions in the membrane |
| $D_{Pt}$ | $cm^2 \cdot s^{-1}$, diffusion coefficient of Pt atoms in the membrane |
| $F$ | $= 96500\ A \cdot s/mol$ , Faraday constant |
| $H_{H_2}$ | dimensionless, Henry constant for hydrogen in membrane |

$H_{O_2}$ dimensionless, Henry constant for oxygen in membrane

$i^0_{H_2}$ $10^{-2}$ *A.cm*$^{-2}$, exchange current for HOR

$i^0_{O_2}$ $10^{-8}$ *A.cm*$^{-2}$, exchange current for ORR

$i_{\lim,O_2}$ *A.cm*$^{-2}$, diffusion limiting current for ORR

$i_{\lim,H_2}$ *A.cm*$^{-2}$, diffusion limiting current for HOR

$J$ *cm*$^{-3}$.*s*$^{-1}$, Average volumetric source of Pt ions

$J_{H_2}$ $mol \cdot (cm^2 \cdot s)^{-1}$, hydrogen diffusion flux to the Pt particle

$J_{O_2}$ $mol \cdot (cm^2 \cdot s)^{-1}$, oxygen diffusion flux to the Pt particle

$k_1$ $1.8 \cdot 10^{-14}$ $mol \cdot (cm^2 \cdot s)^{-1}$, effective reaction rate for Pt dissolution

$k_{\lim Pt^{++}}$ $mol \cdot (cm^2 \cdot s)^{-1}$, diffusion limiting reaction rate for Pt dissolution

$K_{agg}$ *cm*$^{3}$.*s*$^{-1}$, Pt aggregation rate constant

$n_{Pt}$ *cm*$^{-3}$, concentration of Pt particles in the membrane

$N_A$ 6.02 $10^{23}$ *mol*$^{-1}$, Avagadro constant

P *Pa,* pressure

$R$ $= 8.3\ J/mol \cdot \mathrm{K}$, gas constant

$t$ *s*, time

$\tau_{diff}$ *s,* Pt atom diffusion time constant in the membrane

$\tau_{agg}$ *s,* platinum aggregation time constant

$T$ *K*, temperature

$U_0$ 1.23 *V*, equilibrium potential for ORR reaction

$U_1$ 1.188 *V*, equilibrium potential for Pt dissolution reaction

## APPENDIX B

### *Critical Pt particle size for electrochemical reactions*

The electrochemical activity of the platinum catalyst decreases significantly when the particle size is below 1nm [3]. Electrochemical reactions are not energetically favorable on a very small Pt particle, containing just a few Pt atoms, because a charge transfer leads to high charge density on the Pt particle. Thus, there is some critical size, $r_{c,el}$ , of Pt particles, when electrochemical reactions are effective, its estimation is described next.

The rate of electrochemical reactions on Pt particle $K_x$ has exponential dependence on Pt particle potential

$$K_x \sim \exp[\Phi/b]. \quad \text{(A1)}$$

In the case of very small Pt particles, the charge transfer in the electrochemical reactions results in a substantial shift of the Pt particle potential (~*b*), and the electrochemical reactions are suppressed. The dependence of the Pt particle potential on the size $r_{c,el}$ and the charge of the particle can be estimated assuming constant differential capacitance of the Pt surface. The measured value of the differential capacitance is $C_{dif} \approx 30 \mu F/cm^2$ [19].

The critical Pt particle radius is estimated assuming one electron shifts Pt particle potential by a value of $b=0.028V$. Therefore,

$$r_{c,el} = \sqrt{\frac{e}{4\pi b C_{dif}}} \approx 1nm, \quad \text{(A2)}$$

where $e = 1.6 \cdot 10^{-19} C$ is electron charge. As Eq. (A2) is by the order of magnitude estimate of critical radius we assume in the model that critical radius is 2.5nm and coincides with radius of water clusters in hydrated Nafion membrane. In this case Pt particles in membrane simultaneously become electrochemically active and immobile. That assumption simplifies the model without generality loss.

## APPENDIX C

The criterion for diffusion controlled HOR at Pt particle follows from Eq. (29),

$$i_{\lim,H_2} << i^0_{H_2} \exp\left(\frac{\Phi}{b}\right) . \tag{C1}$$

In this case Pt particle disturbs hydrogen concentration in the vicinity of the particle. Combining Eq. (30) for limiting current and Eq. (C1), the particle size when HOR is diffusion controlled can be obtained as

$$r >> 2FC^{ref}_{gas} \frac{H_{H_2} D_{H_2}}{i^0_{H_2} \exp(\Phi/b)} . \tag{C2}$$

Here, $C^{ref}_{gas} = 3.5 \cdot 10^{-5}\, mol/cm^3$ , $H_{H_2} D_{H_2} = 3.6 \cdot 10^{-7}\, cm^2/s$ [22], $i^0_{H_2} = 10^{-2}\, A/cm^2$ [20], $b = RT/F = 0.028V$ . The analogous criterion can be written for ORR. Using mixed potential $\Phi = 0.4V$ in Eq. (C2), see Fig. 2, for Pt particles in the membrane the critical radius when diffusion control starts is estimated at $r = 10^{-10}\, cm$ . As this radius is smaller than atomic radius, the Pt particles of any size perturb the reactants concentration in its vicinity. Therefore, ORR is diffusion controlled in membrane region between $X_0$ and the anode and HOR is diffusion controlled between $X_0$ and cathode. However, it is important to note that these electrochemical reactions are not stable due to reasons discussed in Appendix B.

**APPENDIX D**

In a reaction-diffusion system, the criterion for linear profile of reactant concentration (hydrogen or oxygen) in the membrane is $\Delta J / J << 1$, where J is reactant flux into membrane and $\Delta J$ is reactant consumption in the membrane. Hydrogen flux into membrane is estimated as:

$$J = C_{gas}^{chan} \frac{H_{H_2} D_{H_2}}{L_m} \qquad \text{(D1)}$$

Hydrogen consumption in membrane at Pt particles is

$$\Delta J = \int_0^{L_m} K_{H_2}(x) dx . \qquad \text{(D2)}$$

Substituting reaction rate $K_{H_2}$, Eq. (28), into Eq. (D2) can be written as,

$$\Delta J = \frac{4\pi n_{Pt}}{2F} \int_0^{L_m} r^2(x,t) \cdot i_{H_2}(x,t) dx . \qquad \text{(D3)}$$

As HOR between $X_0$ and the cathode is diffusion controlled, $i_{H_2} = i_{\lim, H_2}$. Assuming uniform Pt particle radius in the membrane and substituting Eq. (30) into Eq. (D3), for limiting current, one obtains

$$\Delta J \approx 4\pi n_{Pt} C_{gas}^{chan} H_{H_2} D_{H_2} L_m (1 - x_0). \qquad \text{(D4)}$$

Using Eq. (D1) and (D4) the criterion for linear hydrogen concentration profile in the membrane is determined as

$$\frac{\Delta J}{J} \approx 4\pi n_{Pt} r L_m^2 << 1 . \qquad \text{(D5)}$$

Analogous estimate can be obtained for oxygen. Using the parameter values of $n_{Pt} = 10^{13} cm^{-3}$, $r = 2.5 nm$ (onset of electrochemical reactions) and $L_m = 20 \mu m$, $4\pi n_{Pt} r L_m^2 \cong 100$. This estimate shows that $O_2$ and $H_2$ concentrations profiles are non-linear by the onset of electrochemical reactions.

## References

1. R. M. Darling and J. P. Meyers, J. Electrochem. Soc., 150(11) (2003) A1523-A1527.
2. R.M. Darling and J.P. Meyers, J. Electrochemical Soc., 152(1) (2005) A242-A247.
3. H. A. Gasteiger, S. Kocha, B. Sompalli and F. T. Wagner, Applied Catalysis B: Environmental Volume 56, Issues 1-2 (2005) 9-35.
4. P. J. Ferreira, G. J. la O', Y. Shao-Horn, D. Morgan, R. Makharia, S. Kocha and H. A. Gasteiger, J. Electrochem. Soc., 152 (2005) A2256-A2271.
5. T. Madden, V. Atrazhev, V. Sultanov, E. Timokhina, S. F. Burlatsky, and M. Gummalla, 211th ECS Meeting, 2007, May 6-10, Chicago, Illinois, Meet. Abstract, Electrochem. Soc. 701, 107 (2007).
6. T. Madden, D. Weiss, N. Cipollini, D. Condit, M. Gummalla, S. Burlatsky, and V. Atrazhev, J. Electrochem. Soc. 156, B657 (2009)
7. M. Gummalla, S. Burlatsky, T. Madden, V. Atrazhev, D, Condit, N. Cipollini, N. Kuzminh, D. Weiss, “Degradation of Polymer-Electrolyte Membranes in Fuel Cells II. Theoretical Model”, J. Electrochem. Soc., submitted
8. Borup et. al., Chem. Rev., 2007, 107 (10), pp 3904–3951
9. T. Patterson, in Fuel Cell Technology Topical Conference Proceedings, AIChE Spring National Meeting, 10-14 March, 2002.

10. K. Yasuda, A. Taniguchi, T. Akita, T. Ioroi and Z. Siroma, Phys. Chem. Chem. Phys., 8 (2006) 746-752.
11. J. Zhang, B. A. Litteer, W. Gu, H. Liu, and H. A. Gasteiger, J. Electrochem Soc, 154, 10, (2007) B1006-B1011
12. J. Péron, Y. Nedellec, D. J. Jones, J. Rozière, Journal of Power Sources 185 (2008) 1209–1217
13. W. Bi, G. E. Gray, and T. F. Fuller, Electrochem. Solid-State Lett., Volume 10, Issue 5, pp. B101-B104 (2007)
14. A. Ohma, S. Suga, S. Yamamoto, and K. Shinohara, J. Electrochem. Soc., Volume 154, Issue 8, pp. B757-B760 (2007)
15. V. V. Atrazhev, N.S. Erikhman, S.F. Burlatsky, J. Electroanal. Chem. 601 (2007) 251–259
16. W. Liu and D. Zuckerbrod, J. Electrochem. Soc. 152(6) (2005) A1165-A1170.
17. F.Leyvraz, Physics Reports 383 (2003) 95–212.
18. P. G. J. van Dongen, Phys. Rev. Lett., 63 (12), (1989), 1281-1284.
19. K. J. Vetter, *Electrochemische Kinetik*, Springer-Verlag, Berlin etc., 1961
20. P. Gode et al, J. Electroanal. Chem. 518 (2002) 115–122

**List of Figures**

Figure 1: Pt particle size in the membrane along its length, from TEM images of a) $H_2$/Air cycled cell and b) $H_2/O_2$ cycled cell.

Figure 2: The potential of membrane in an MEA as a function of the platinum probe size. The solid lines are model predictions and the blue data points are experimental probe potentials reported in [16].

Figure 3: Schematic that shows the mechanism of the platinum dissolution at high potentials and precipitation at low potentials, within the membrane.

Figure 4: Schematic representation of hydrogen and oxygen concentration profiles in the membrane in the a) nucleation stage, b) growth stage.

Figure 5: The model predicted profiles of oxygen, hydrogen, Pt particles and potential in the membrane for pure hydrogen and oxygen at the anode and cathode respectively, at a) t=0.1 hr, b) t=10 hrs, c) t=100 hrs.

Figure 6: SEM image of membrane-electrode assembly after potential cycling under $H_2$/air conditions [9]; b) model predicted evolution of Pt particle potential in membrane; c) model predicted evolution of Pt particle radius in the membrane

Table 1. Model equations and boundary conditions.

| Notation | Description | Equation | Boundary conditions |
|---|---|---|---|
| $C_{O_2}(x,t)$ | $O_2$ concentration, mol/cm$^3$ | $D_{O_2}\frac{\partial^2 C_{O_2}}{\partial x^2}=K_{O_2}$ | $C_{O_2}\big\vert_{AMI}=0$ <br> $C_{O_2}\big\vert_{CMI}=H_{O_2}C_{O_2}^{chan}$ |
| $C_{H_2}(x,t)$ | $H_2$ concentration, mol/cm$^3$ | $D_{H_2}\frac{\partial^2 C_{H_2}}{\partial x^2}=K_{H_2}+K_{Pt^{++}H_2}$ | $C_{H_2}\big\vert_{CMI}=0$ <br> $C_{H_2}\big\vert_{AMI}=H_{H_2}C_{H_2}^{chan}$ |
| $C_{Pt^{++}}(x,t)$ | Concentration of $Pt^{++}$ ions in solution, mol/cm$^3$ | $D_{Pt^{++}}\frac{\partial^2 C_{Pt^{++}}}{\partial x^2}=-K_{Pt^{++}}+K_{Pt^{++}H_2}$ | $-D_{Pt^{++}}\frac{dC_{Pt^{++}}}{dx}\Big\vert_{CMI}=J_{Pt,total}$ <br> $C_{Pt^{++}}\big\vert_{AMI}=0$ |
| $C_{Pt}(x,t)$ | Concentration of free Pt atoms in solution, mol/cm$^3$ | $D_{Pt}\frac{\partial^2 C_{Pt}}{\partial x^2}=-K_{Pt^{++}H_2}+K_{Pt}$ | $C_{Pt}\big\vert_{CMI}=0$ <br> $C_{Pt}\big\vert_{AMI}=0$ |
| $\Phi(x,t)$ | Electric potential of Pt particle, V | $2K_{O_2}-K_{H_2}-K_{Pt^{++}}=0$ | |
| $r(x,t)$ | Radius of Pt particles, cm | $\frac{4\pi r^2 n_{Pt}}{v_{Pt}}\frac{\partial r(x,t)}{\partial t}=K_{Pt}-K_{Pt^{++}}$ | Initial condition <br> r(x,0) = 3 nm |

Table 2. Reactions incorporated in the model.

| Reaction | Reaction rate, $\left[\frac{mol}{cm^3 \cdot s}\right]$ |
|---|---|
| ORR (oxygen reduction reaction)<br>$O_2 + 4H^+ + 4e^- \leftrightarrow 2H_2O$ | $K_{O_2}(x,t) = \frac{4\pi r^2(x,t) n_{Pt} i^0_{O_2}}{4F} \cdot \frac{\frac{C_{O_2}}{H_{O_2} C^{ref}_{gas}} \exp\left(\frac{U_0 - \Phi}{b}\right) - \exp\left(-\frac{U_0 - \Phi}{b}\right)}{1 + \frac{i^0_{O_2}}{i_{\lim, O_2}} \exp\left(\frac{U_0 - \Phi}{b}\right)}$ |
| HOR (hydrogen oxidation reaction)<br>$H_2 \leftrightarrow 2H^+ + 2e^-$ | $K_{H_2}(x,t) = \frac{4\pi r^2(x,t) n_{Pt} i^0_{H_2}}{2F} \cdot \frac{\frac{C_{H_2}}{H_{H_2} C^{ref}_{gas}} \exp\left(\frac{\Phi}{b}\right) - \exp\left(-\frac{\Phi}{b}\right)}{1 + \frac{i^0_{H_2}}{i_{\lim, H_2}} \exp\left(\frac{\Phi}{b}\right)}$ |
| Pt dissolution / precipitation<br>$Pt \leftrightarrow Pt^{++} + 2e^-$ | $K_{Pt^{++}} = 4\pi r^2 n_{Pt} k_1 \frac{\exp\left(2\frac{\alpha_1(\Phi - U_1)}{b}\right) - \frac{C_{Pt^{++}}}{C^{ref}_{ion}} \exp\left(-2\frac{(1-\alpha_1)(\Phi - U_1)}{b}\right)}{1 + \frac{k_1}{k_{\lim Pt^{++}}} \exp\left(-2\frac{(1-\alpha_1)(\Phi - U_1)}{b}\right)}$ |
| Pt ions reduction by hydrogen<br>$Pt^{++} + H_2 \rightarrow Pt^0 + 2H^+$ | $K_{Pt^{++}H_2} = 4\pi D_{H_2} a_{Pt} N_A C_{H_2} C_{Pt^{++}}$ |
| Aggregation of neutral Pt atoms with Pt particles<br>$(Pt)_n + Pt^0 \rightarrow (Pt)_{n+1}$ | $K_{Pt} = 4\pi D_{Pt} r n_{Pt} C_{Pt}$ |

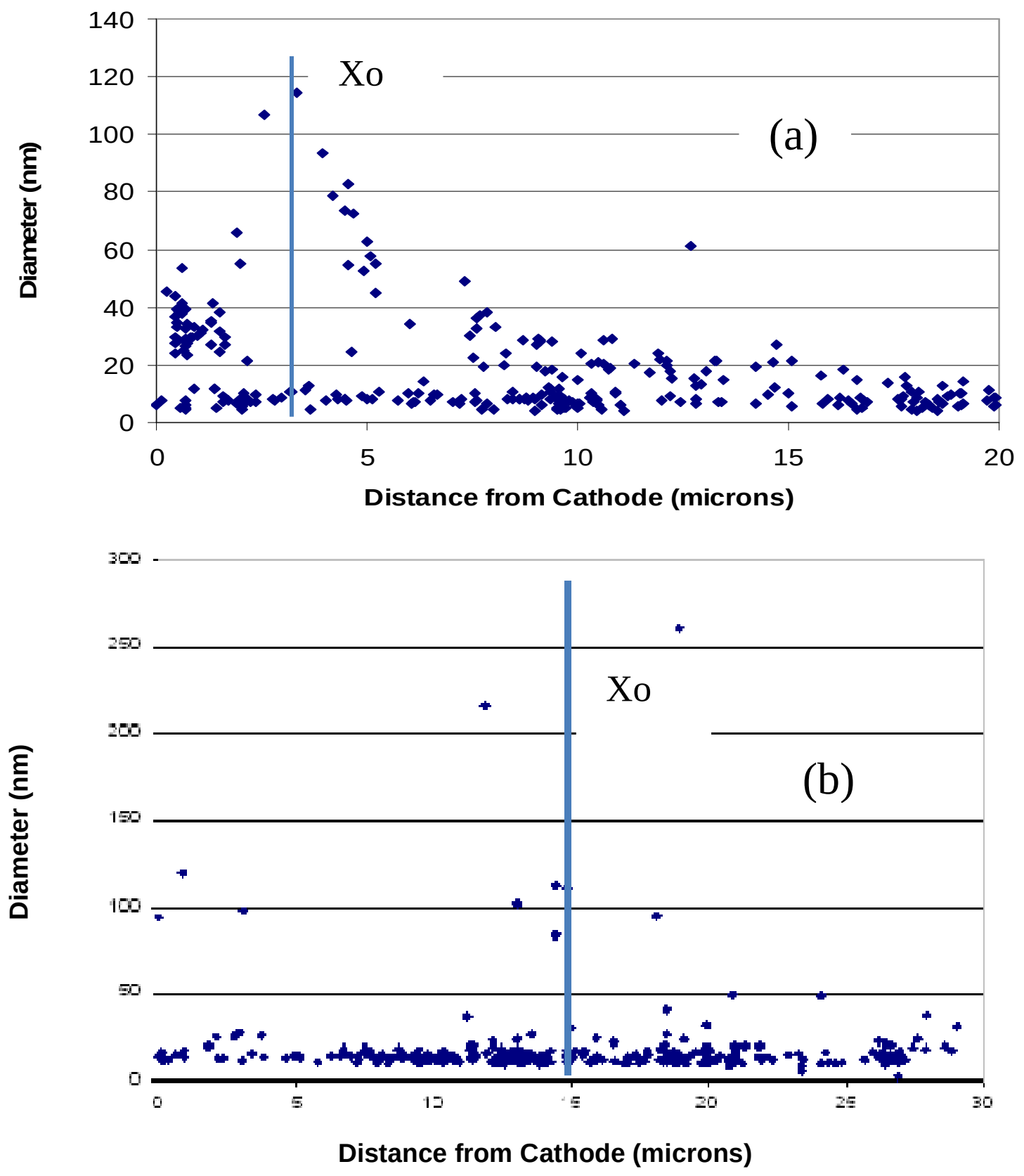

**Figure 1**

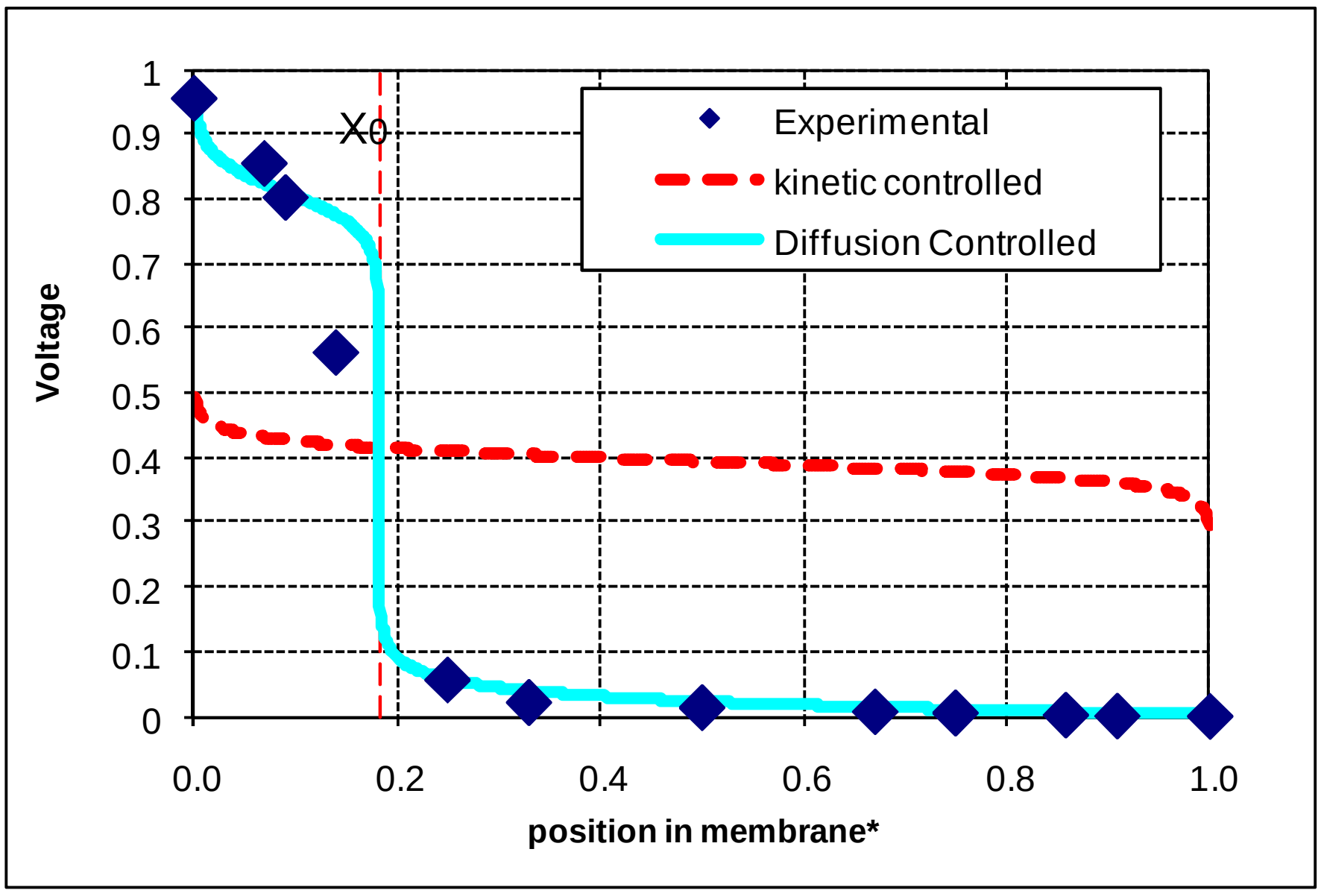

**Figure 2**

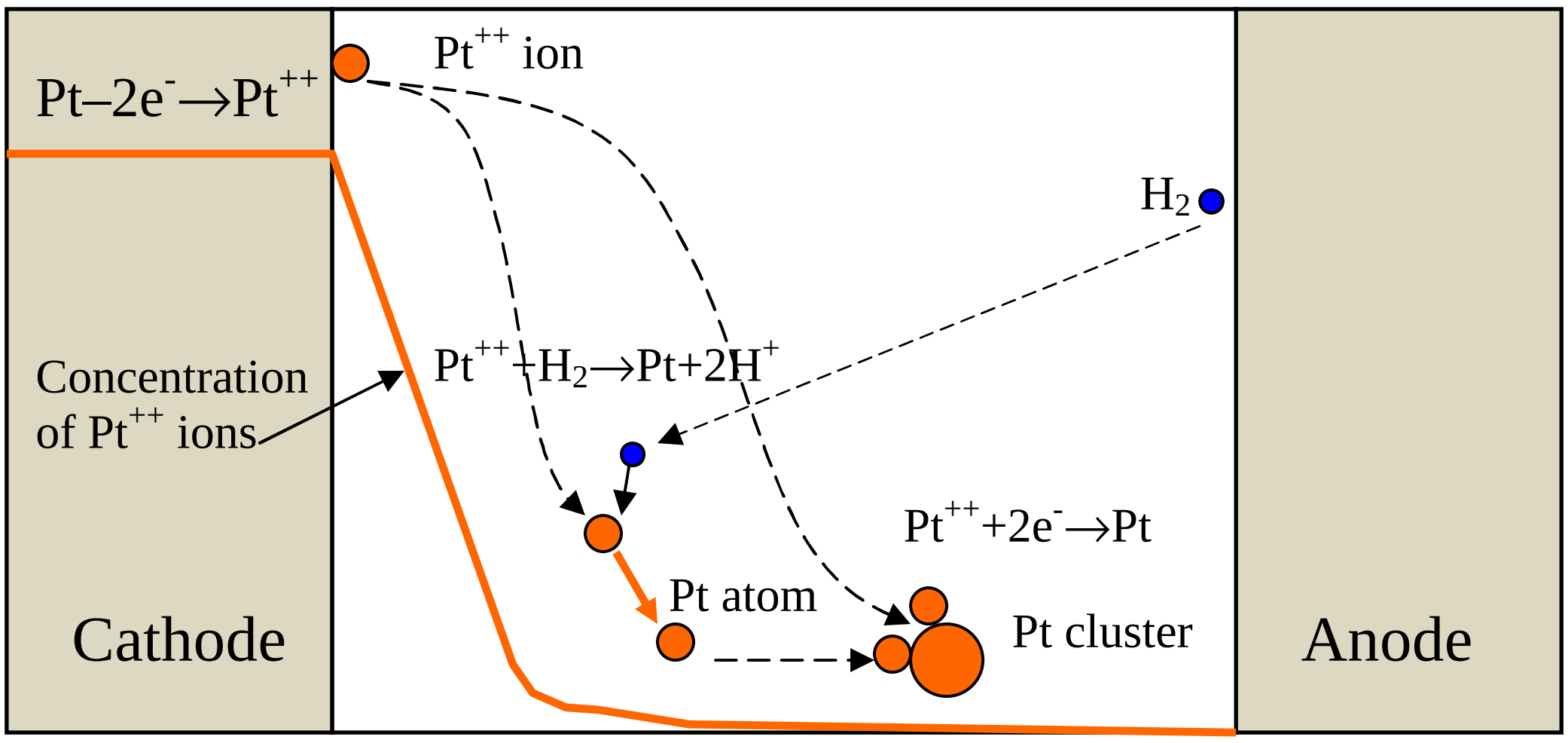

**Figure 3**

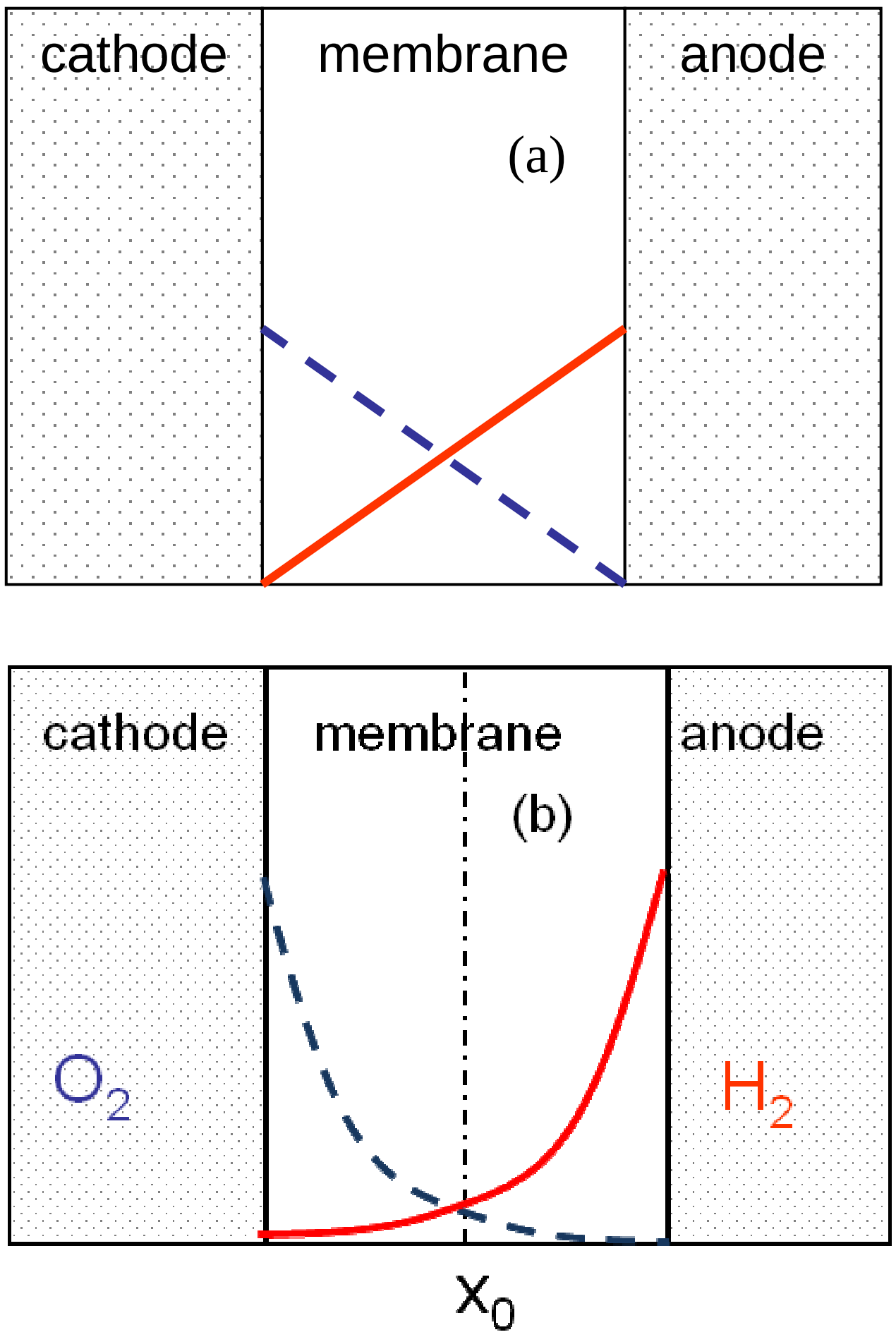

**Figure 4**

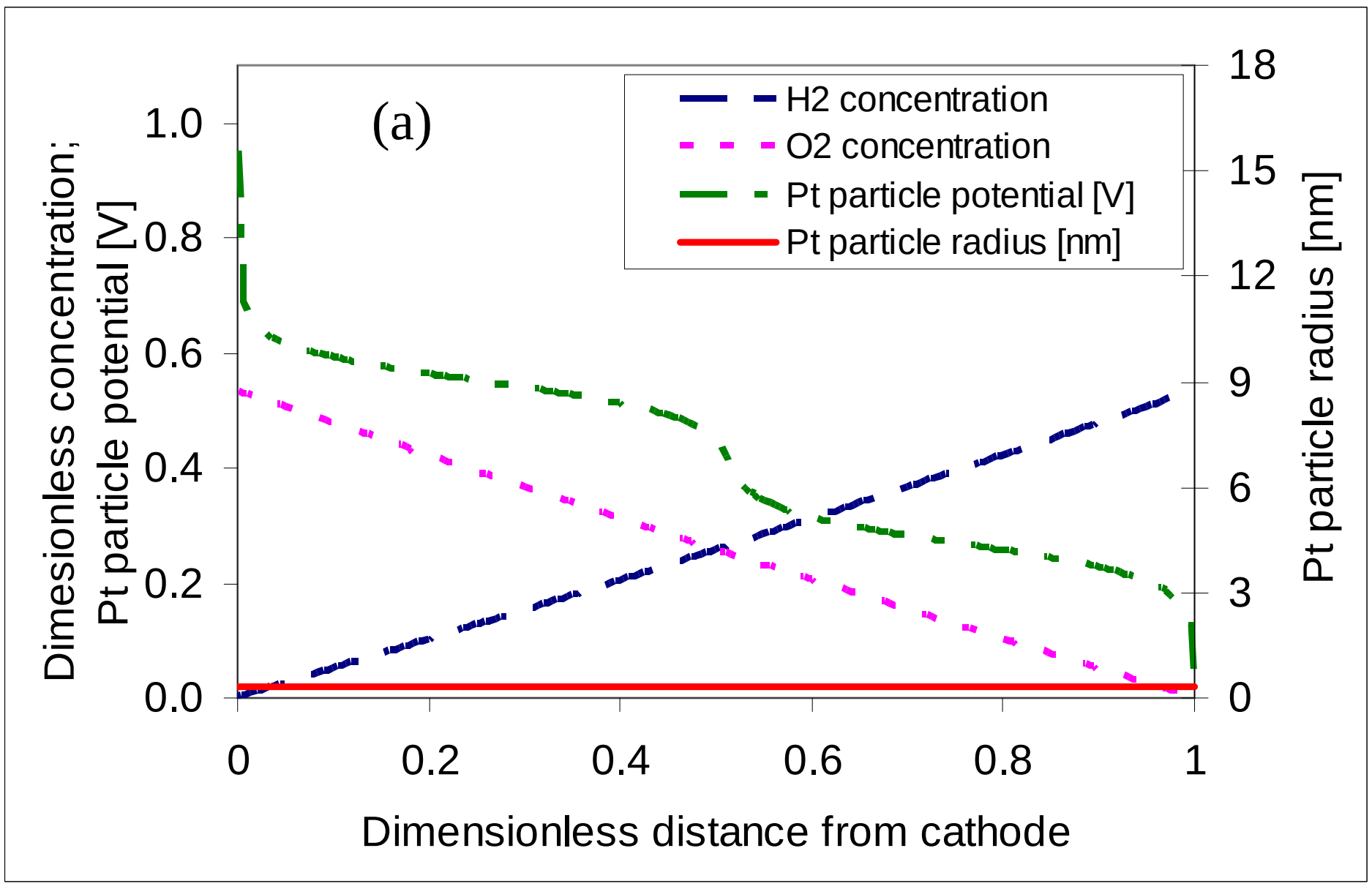

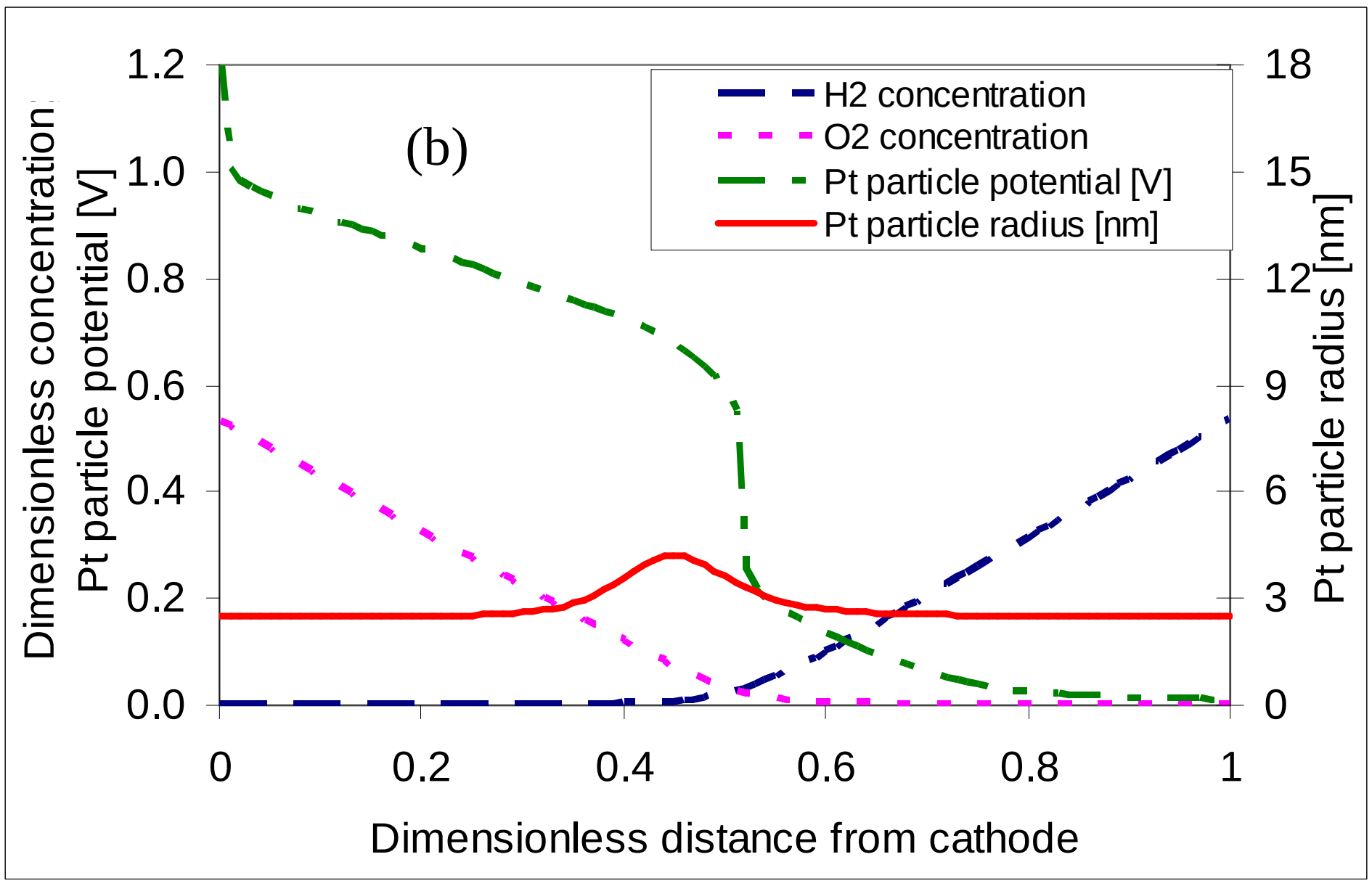

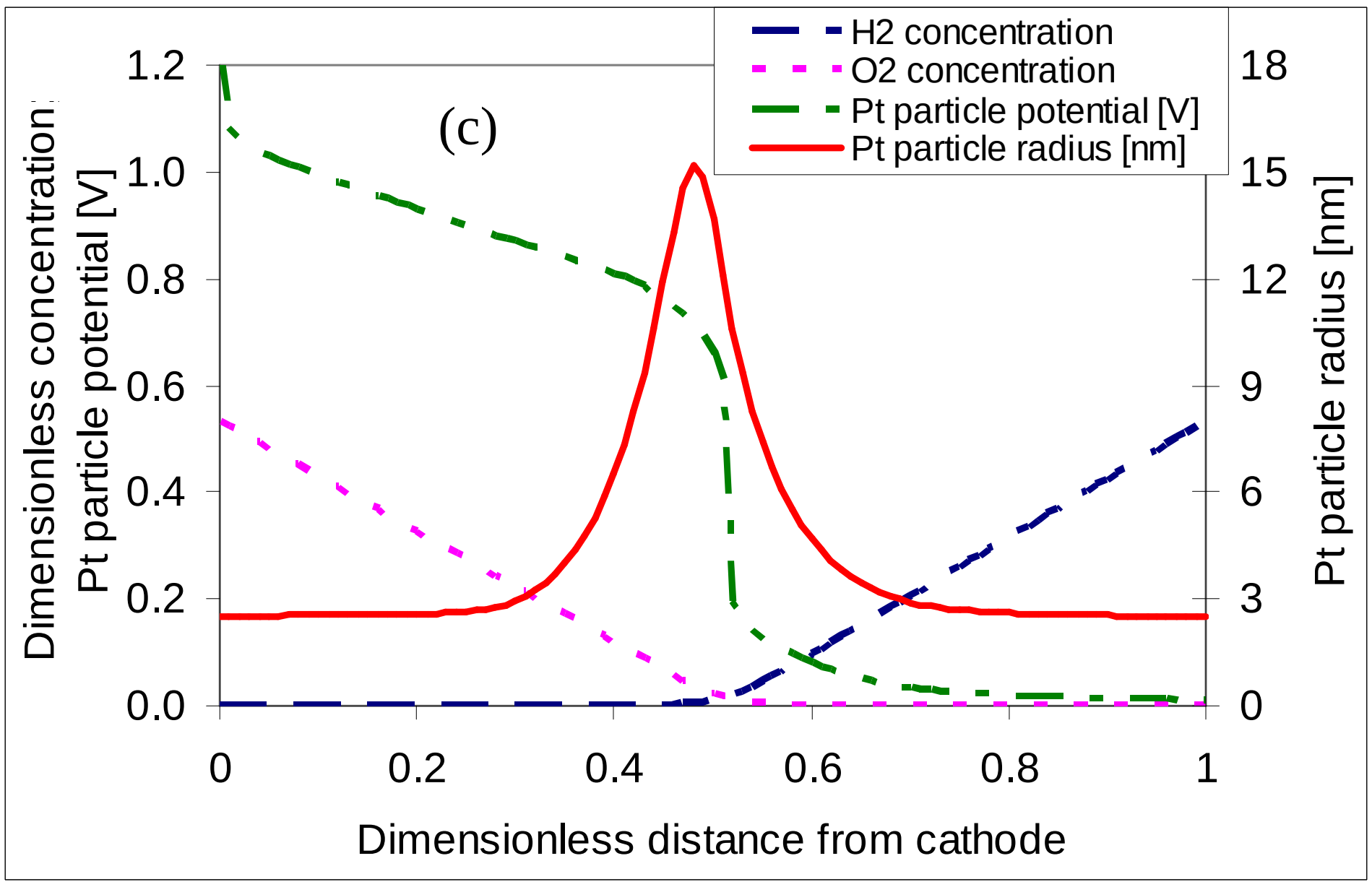

Figure 5

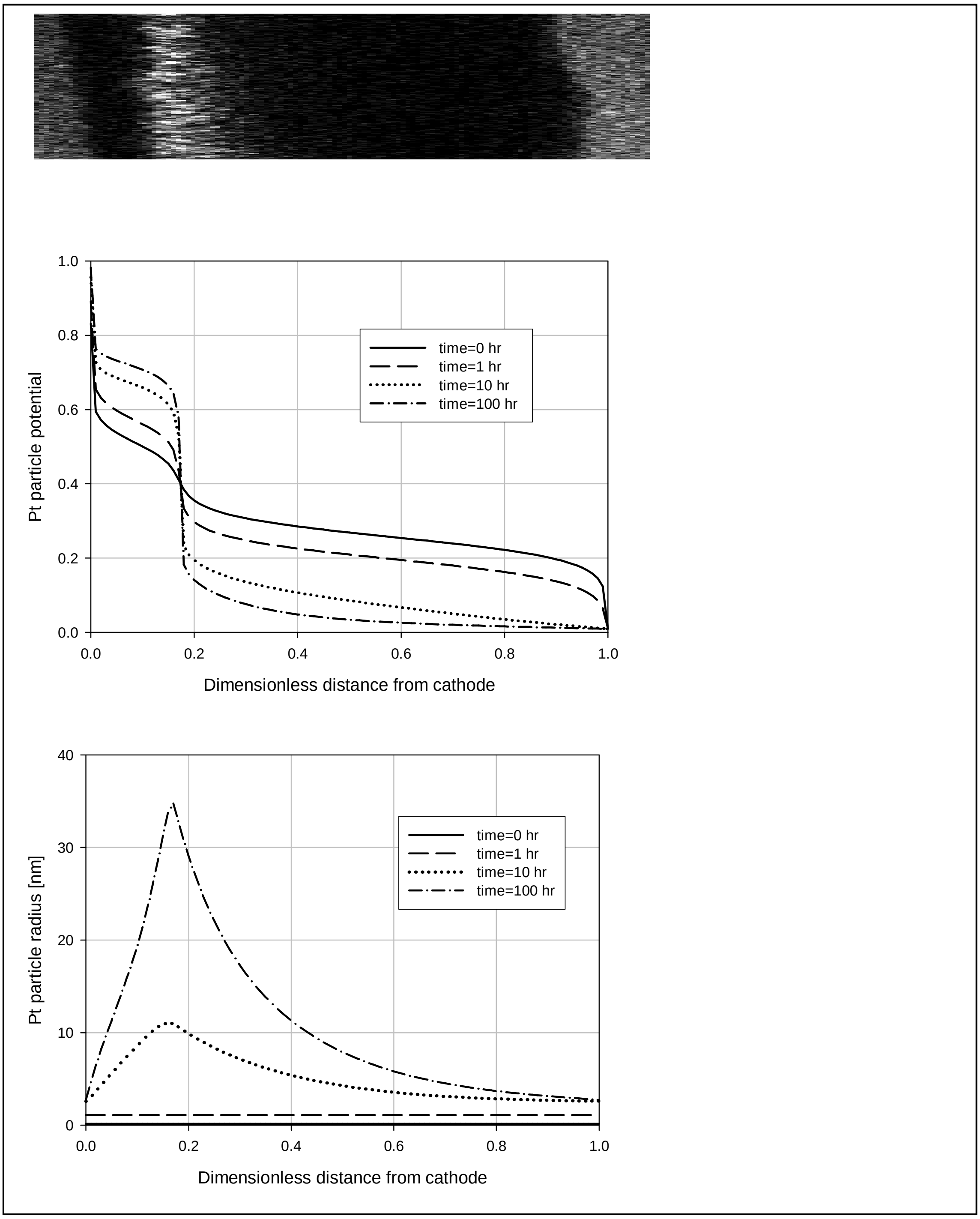

**Figure 6**